\documentclass{elsart5p}

\usepackage{graphicx}
\usepackage{dcolumn}
\usepackage{bm}
\usepackage{amssymb}

\begin{document}

\begin{frontmatter}

\title{Gap-mediated magnetization of a pseudo-one-dimensional system with a spin-orbit interaction}

\author[hatano]{Naomichi Hatano}
\ead{hatano@iis.u-tokyo.ac.jp}
\author[shirasaki]{Ry\=oen Shirasaki}
\ead{sirasaki@phys.ynu.ac.jp}
\author[nakamura]{Hiroaki Nakamura}
\ead{nakamura@tcsc.nifs.ac.jp}

\address[hatano]{Institute of Industrial Science, University of Tokyo, Komaba, Meguro, Tokyo 153-8505, Japan}
\address[shirasaki]{Department of Physics, Yokohama National University, Tokiwadai, Hodogaya-ku, Yokohama, Kanagawa 240-8501, Japan}
\address[nakamura]{Theory and Computer Simulation Center, National Institute for Fusion Science,
Oroshi-cho, Toki, Gifu 509-5292, Japan}

\date{\today}

\begin{abstract}
We argue that a pseudo-one-dimensional electron gas is magnetized when a voltage bias is applied with the Fermi level tuned to be in the energy gap generated by a spin-orbit interaction.
The magnetization is an indication of spin-carrying currents due to the spin-orbit interaction.
The origin of the magnetization, however, is essentially different from the ``spin accumulation" in two-dimensional systems with spin-orbit interactions.
\end{abstract}

\end{frontmatter}

\section{Introduction}

\begin{figure}[b]
\centering
\includegraphics[width=0.6\columnwidth]{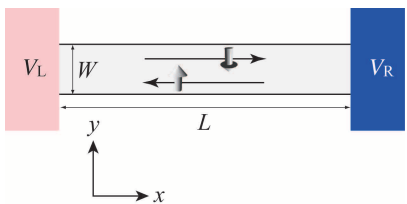}
\caption{(color online) A quantum wire under a voltage gradient.}
\label{fig_system}
\end{figure}

In the present Letter, we consider a quasi-one-dimensional electron gas with the Rashba or Dresselhaus spin-orbit interaction~\cite{Elliott54,Messiah58,Rashba60,Dresselhaus55}.
We argue that the system is magnetized when a voltage bias is applied longitudinally
 (Fig.~\ref{fig_system}) 
with the chemical potential tuned to be in the energy gap generated by the spin-orbit interaction.
The magnetization indicates spin-carrying currents due to the spin-orbit interaction.
The origin of the magnetization, however, is entirely novel;
it is essentially different from the origin of the ``spin accumulation"~\cite{Edelstein90,spinaccumref1,spinaccumref2,spinaccumref3,Kleinert05} in two-dimensional systems with spin-orbit interactions.

\begin{figure}
\centering
\includegraphics[width=0.7\columnwidth]{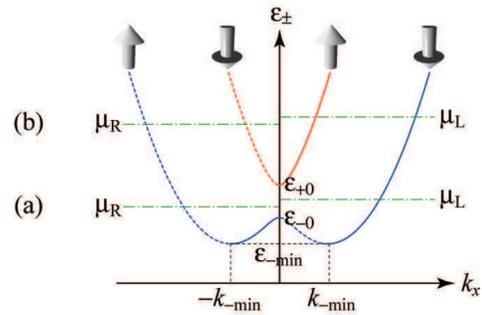}
\caption{(color online) A schematic view of the dispersion relations of the upper and lower subbands of a channel.
The right-going current has a chemical potential $\mu_\mathrm{L}=-|e|V_\mathrm{L}$ of the left contact, while the left-going current has a chemical potential $\mu_\mathrm{R}=-|e|V_\mathrm{R}$ of the right contact.}
\label{fig_channel}
\end{figure}

A spin-orbit interaction affects the dispersion of a conducting channel of a pseudo-one-dimensional system in two ways (Fig.~\ref{fig_channel}):
first, the dispersion of up-spin electrons and that of down-spin electrons, respectively, shift sideways in the opposite directions;
next, the crossing of the dispersions at $k_x=0$ opens up an energy gap (see, however, the discussions in \S4).
Under this dispersion, we consider a simple theoretical state called the non-equilibrium steady state~\cite{Ruelle00,Ogata02a,Ogata02b};
the right-going current has the Fermi distribution of the left contact while the left-going current has the Fermi distribution of the right contact and they run ballistically and independently without reflection.
When the chemical potentials (the Fermi levels) of the right and left contacts are tuned to be in the energy gap of the dispersion (the case (a) in Fig.~\ref{fig_channel}), the number of up-spin electrons in the left-going current and the number of down-spin electrons in the right-going current can differ, and thereby appears the magnetization.
This argument does not apply when the chemical potential is in the middle of a subband (the case (b) in Fig.~\ref{fig_channel}), where the right-going up-spin electrons cancel the magnetization of the right-going down-spin electrons.

Magnetic properties of mesoscopic systems with spin-orbit interactions are of great interest recently, particularly from the viewpoint of spintronics~\cite{Zutic04,Murakami03,Eto05};
we could control the dynamics of spins with an external electric field.
Most studies so far have focused on two-dimensional systems.
The spin accumulation~\cite{Edelstein90,spinaccumref1,spinaccumref2,spinaccumref3,Kleinert05} and the spin-Hall effect~\cite{Murakami03,spinHallref1,spinHallref2,spinHallref3,Sugimoto06} are two of the phenomena much discussed in two dimensions.
The recent study by Eto \textit{et al.}~\cite{Eto05} showed the occurrence of a spin-polarized current in a pseudo-one-dimensional system with the Rashba spin-orbit interaction.
This let us realize that pseudo-one-dimensional systems also contain much interesting physics of spin-orbit interactions.

We may say that the origin of exotic phenomena due to spin-orbit interactions in pseudo-one-dimensional systems is essentially different from the one in two-dimensional systems.
Both the spin-polarized current predicted by Eto \textit{et al.}~\cite{Eto05} and the magnetization predicted here are due to an avoided crossing in the dispersion of the pseudo-one-dimensional system.
We also note that we can treat pseudo-one-dimensional systems much simpler than two-dimensional systems.
In fact, the linear response of the spin Hall effect~\cite{Murakami03,spinHallref1} is recently under debate;
it was shown~\cite{spinHallref2,spinHallref3} that the spin Hall coefficient vanishes in the DC limit $\omega\to 0$ of the electric field under the presence of impurity scattering.
It was then argued recently~\cite{Sugimoto06} that the long-range impurities may not eliminate the spin Hall coefficient even in the DC limit.
In contrast, the treatment of the electronic transport in one dimension is well established.
Our argument here is based on a common method widely used in studies of electronic transport in one dimension~\cite{Datta}.
We have already confirmed that inclusion of elastic impurities does not alter our conclusion qualitatively~\cite{Shirasaki_note}.

\section{Analytic calculation for a continuum model}
\label{sec-analytic}

Let us begin our explanation with the Rashba system.
The treatment of the Dresselhaus system is not much different.
The Hamiltonian of the system with the Rashba spin-orbit interaction is given by~\cite{Rashba03}
\begin{equation}
\label{eq10}
\hat{\mathcal{H}}=
\frac{{\hat{p}_x}^2+{\hat{p}_y}^2}{2m^\ast}
+\alpha_\mathrm{RSO} \left(
\hat{p}_x \hat{\sigma}_y
-\hat{p}_y \hat{\sigma}_x
\right),
\end{equation}
where $m^\ast$ denotes the effective mass of an electron and $\alpha_\mathrm{RSO}$ denotes the strength of the Rashba interaction.
From a simplistic point of view, we could say that the term $\hat{p}_y\hat{\sigma}_x$ generates the gap at $k_x=0$ while the term $\hat{p}_x\hat{\sigma}_y$ helps the current in the $x$ direction to induce the magnetization in the $y$ direction.

We choose the $y$ direction as the quantization axis of the electron spin.
 and hence use the following representations hereafter:
\begin{equation}
\label{eq20}
\hat{\sigma}_x=\left(
\begin{array}{cc}
0 & 1 \\
1 & 0
\end{array}
\right),
\quad
\label{eq30}
\hat{\sigma}_y=\left(
\begin{array}{cc}
1 & 0 \\
0 & -1
\end{array}
\right),
\quad
\label{eq40}
\hat{\sigma}_z=\left(
\begin{array}{cc}
0 & i \\
-i & 0
\end{array}
\right).
\end{equation}
We assume that the system has only one mode in the $z$ direction.
The system is of length $L$ in the $x$ direction and of width $W$ in the $y$ direction with $W\ll L\ll\mbox{mean free path}$.
Hence the electrons run along the wire ballistically through a few channels.

\begin{figure}
\centering
\includegraphics[width=0.8\columnwidth]{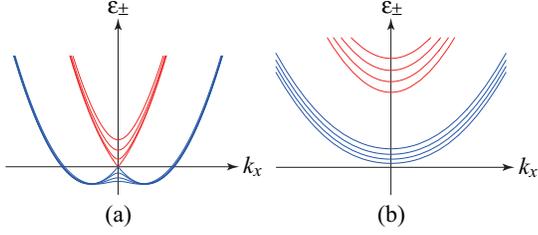}
%
%
%
%
%
\caption{(color online) Schematic views of the dispersion relation~(\ref{eq70}) for (a) some of the lowest channels and for (b) higher channels.
The lowest channel has a real crossing at $k_x=0$, but a higher channel has an avoided crossing.
In (b), the double minima of the lower branch vanish for a high channel and the whole dispersion shifts upwards as we go to even higher channels.}
\label{fig_disp}
\end{figure}

We first diagonalize the Hamiltonian~(\ref{eq10}) in the momentum space with the bases
\begin{equation}
\label{eq50}
\left| k_x, k_y, \sigma_y \right\rangle
=\frac{1}{\sqrt{LW}} e^{i(k_x x+k_yy)} \left|\sigma_y\right\rangle,
\end{equation}
where $k_y=0, \pm 2\pi/W, \pm 4\pi/W,\ldots$ and $\sigma_y=\uparrow,\downarrow$.
(For simplicity, we here imposed periodic boundary conditions in the $y$ direction,
which might be indeed materialized as a nanotube~\cite{nanotube}.
See the discussions in \S4 below for fixed boundary conditions.)
The Hamiltonian~(\ref{eq10}) is given by a two-by-two matrix in the spin space in the form
\begin{equation}
\label{eq60}
\hat{\mathcal{H}}=\frac{\hbar^2}{2m^\ast}\left[k^2+\theta k\left(\hat{\sigma}_y\cos\phi-\hat{\sigma}_x\sin\phi\right)\right],
\end{equation}
where $k^2\equiv {k_x}^2+{k_y}^2$, $(k_x,k_y)=k(\cos\phi,\sin\phi)$ and $\theta\equiv 2m^\ast\alpha_\mathrm{RSO}/\hbar$.
The spin rotation $\hat{U}\equiv\exp(i\phi\hat{\sigma}_z/2)$ gives the representation
\begin{equation}
\label{eq80}
\hat{U}\hat{\mathcal{H}}\hat{U}^\dag=\frac{\hbar^2}{2m^\ast}\left(k^2+\theta k\hat{\sigma}_y\right)
\end{equation}
with the eigenvalues~\cite{Molenkamp01,Streda03}
\begin{equation}
\label{eq70}
\varepsilon_\pm (k_x,k_y)=\frac{\hbar^2}{2m^\ast}\left(k^2\pm\theta k\right),
\end{equation}
where the positive sign denotes the upper subband and the negative sign denotes the lower subband.
For the lowest channel $k_y=0$, the dispersion has two branches of parabolas with up and down spins, respectively.
In the higher channels $k_y>0$, the two branches are mixed around the avoided crossing at $k_x=0$ (Fig.~\ref{fig_disp}(a)).
Note that, away from the avoided crossing, the lower branch in the region $k_x\gg 0$ still predominantly has down spins while it almost has up spins in the region $k_x\ll 0$.

Incidentally, lower channels are nearly degenerate in the present system as shown in Fig.~\ref{fig_disp}(a).
This near degeneracy may be lifted by a confining potential $U(y)$.
As another remark, the double minima of the lower branch shown in Fig.~\ref{fig_disp}(a) vanish in higher channels and the dispersion shifts upwards in even higher channels (Fig.~\ref{fig_disp}(b)).
The algebra hereafter slightly changes in the latter case but the final result~(\ref{eq180}) below is still valid.

The right-going current originated in the left contact with the chemical potential $\mu_\mathrm{L}$ contains all the states with positive group velocities.
The group velocity in the $x$ direction is
\begin{equation}
\label{eq100}
v_\pm=\frac{1}{\hbar}\frac{\partial\varepsilon_\pm(k_x,k_y)}{\partial k_x},
\end{equation}
which is positive wherever the slope of the dispersion is positive (the solid curves in Fig.~\ref{fig_channel}).
Each state $(k_x,k_y,\pm)$ has the magnetization per unit area in the $y$ direction of the form
\begin{equation}
\label{eq110}
m_\pm(k_x,k_y)=\pm\frac{\mu_\mathrm{e}}{LW}\cos\phi=\pm\frac{\mu_\mathrm{e}}{LW}\frac{k_x}{k}
\end{equation}
because the spin rotation $\hat{U}$ changes the spin operator $\hat{\sigma}_y$ to $\hat{\sigma}_y\cos\phi+\hat{\sigma}_x\sin\phi$.
Thus we have the magnetization per unit area of the right-going current as
\begin{eqnarray*}
m_\mathrm{R}(k_y)=
\left(\int_{-k_{-\mathrm{min}}}^0+\int_{k_{-\mathrm{min}}}^\infty\right)
m_-f(\varepsilon_-;T,\mu_\mathrm{L})\frac{dk_x}{2\pi/L}
\nonumber\\
+\int_0^\infty m_+f(\varepsilon_+;T,\mu_\mathrm{L})\frac{dk_x}{2\pi/L}
\end{eqnarray*}
\vspace*{-1.5\baselineskip}
\begin{equation}
\label{eq170}
=\frac{\mu_\mathrm{e}m^\ast}{\pi\hbar^2 W}\left(
\int_{\varepsilon_{-\mathrm{min}}}^{\varepsilon_{-0}}
-\int_{\varepsilon_{-\mathrm{min}}}^\infty
+\int_{\varepsilon_{+0}}^\infty
\right)
\frac{f(\varepsilon;T,\mu_\mathrm{L})}{\sqrt{\theta^2+\frac{8m^\ast}{\hbar^2}\varepsilon}}d\varepsilon
\end{equation}
for a channel, where $f(\varepsilon;T,\mu)$ denotes the Fermi distribution with the temperature $T$ and the chemical potential $\mu$ and we defined for each channel the following variables:
\begin{eqnarray}
&&k_{-\mathrm{min}}(k_y)\equiv\sqrt{\theta^2/4-{k_y}^2},
\qquad
\varepsilon_{-\mathrm{min}}\equiv-\frac{\hbar^2\theta^2}{8m^\ast},
\nonumber\\
\label{eq140}
&&\varepsilon_{\pm 0} (k_y) \equiv \frac{\hbar^2}{2m^\ast}\left({k_y}^2\pm \theta k_y\right);
\end{eqnarray} 
see Fig.~\ref{fig_channel}.
In changing the integration variable from $k_x$ to $\varepsilon$ in Eq.~(\ref{eq170}), we used the density of states $D_\pm$ 
given by
\begin{equation}
\label{eq90}
\frac{1}{D_\pm}=\frac{2\pi}{L} \left|\frac{\partial\varepsilon_\pm}{\partial k_x}\right|
=\frac{\pi\hbar^2}{m^\ast L}\frac{\left|k_x\right|}{k}\sqrt{\theta^2+\frac{8m^\ast}{\hbar^2}\varepsilon_\pm}.
\end{equation}
The van Hove singularity is canceled in Eq.~(\ref{eq170}) by the zero of $m_\pm$.
We can likewise obtain the magnetization $m_\mathrm{L}(k_y)$ of the left-going current.
The total magnetization per unit area is given by
\begin{eqnarray}
\lefteqn{
M_\mathrm{tot}\equiv\sum_{k_y}m_\mathrm{R}(k_y)+m_\mathrm{L}(k_y)
}
\label{eq180}
\nonumber\\
&=&-\frac{\mu_\mathrm{e}}{2\pi\hbar W}\sum_{k_y}
\int_{\varepsilon_{-0}(k_y)}^{\varepsilon_{+0}(k_y)}
\frac{f(\varepsilon;T,\mu_\mathrm{L})-f(\varepsilon;T,\mu_\mathrm{R})}%
{\sqrt{{\alpha_\mathrm{RSO}}^2+2\varepsilon/m^\ast}}d\varepsilon.
\nonumber\\
\end{eqnarray}
Note that only the integration over the energy gap survives.
This is consistent with what we described in Fig.~\ref{fig_channel}.

Let us analyze the linear response.
We define the chemical-potential bias as 
\begin{equation}
\label{eq200}
\mu_\mathrm{L}=\mu+\frac{\Delta \mu}{2}
\quad
\mbox{and}
\quad
\mu_\mathrm{R}=\mu-\frac{\Delta \mu}{2}
\end{equation}
with $\Delta V=V_\mathrm{L}-V_\mathrm{R}=-\Delta\mu/|e|$.
The expansion of Eq.~(\ref{eq180}) with respect to $\Delta\mu$ is followed by
\begin{equation}
\label{eq240}
\frac{M_\mathrm{tot}}{\Delta V}\simeq
\sum_{k_y}M^{(1)}(k_y)\equiv
\frac{\mu_\mathrm{e} |e|}{2\pi\hbar W}\sum_{k_y}B(k_y),
\end{equation}
where
\begin{eqnarray}
\label{eq250}
B(k_y)\equiv
\int_{x_{-0}(k_y)}^{x_{+0}(k_y)}
\frac{g(x)}{\sqrt{{\alpha_\mathrm{RSO}}^2+2(k_\mathrm{B}Tx+\mu)/m^\ast}}dx
\end{eqnarray}
with
\begin{equation}
g(x)\equiv \frac{1}{[2\cosh(x/2)]^{2}}
\quad
\mbox{and}
\quad
x_{\pm 0}(k_y)\equiv \frac{\varepsilon_{\pm 0}(k_y)-\mu}{k_\mathrm{B}T}.
\end{equation}

\section{Numerical demonstration}

\begin{figure}
\centering

\includegraphics[width=0.8\columnwidth]{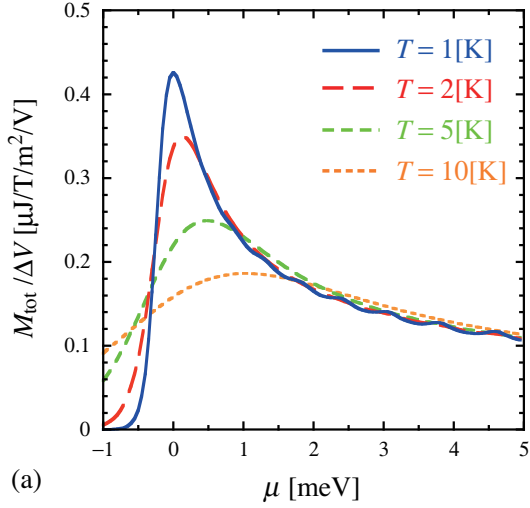}

\vspace{\baselineskip}

\includegraphics[width=0.8\columnwidth]{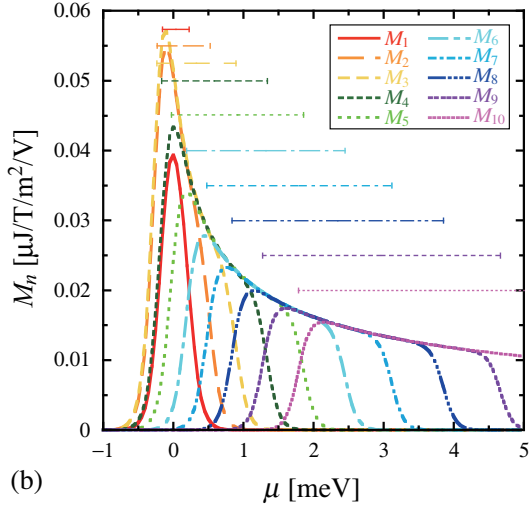}
\caption{(color online) (a) The response of the magnetization per unit area to the voltage bias $\Delta V$.
(b) Contributions $M_n\equiv M^{(1)}(2n\pi/W)$ from various channels for $T=1$[K].
The bars indicate the energy gaps of the channels.
In higher channels, the double minima at $k_x=k_{-\mathrm{min}}$ vanish and the \lq\lq energy gap\rq\rq\ shifts to the right.}
\label{fig_M_V}
\end{figure}

Figure~\ref{fig_M_V} shows the result of numerical calculation, where we used the values for an InGaAs/InAlAs heterojunction~\cite{Sato01}: $\alpha_\mathrm{RSO}\hbar= 3\times 10^{-11}$[eV m] and $m^\ast=0.041 m_e$.
We also set $W=1$[$\mu$m].
We find in Fig.~\ref{fig_M_V}(a) a peak of the magnetization around the energy gaps of the dispersion relation.
We observe for $\Delta V=1$[mV] the magnetization of the order of $10^{-10}$[J/T/m$^2$], which translates to $10^{-5}$[G] if we assume that the thickness of the heterojunction is $10$[nm]~\cite{Sato01}.

Figure~\ref{fig_M_V}(b) shows the contribution of each channel, $M_n\equiv M^{(1)}(2n\pi/W)$, to the total magnetization $M_\mathrm{tot}$ at $T=1$[K].
We note that each contribution is finite over the range of the energy gap of the respective channel, which is indicated by each bar.
In the present case, some of the energy gaps overlap and hence a sharp peak in Fig.~\ref{fig_M_V}(a).
Figure~\ref{fig_M_V}(b) also tells us that the ripples in the tail of the curve for $T=1$[K] in Fig.~\ref{fig_M_V}(a) is due to the summation over the channels.

\begin{figure}
\centering
\includegraphics[width=0.8\columnwidth]{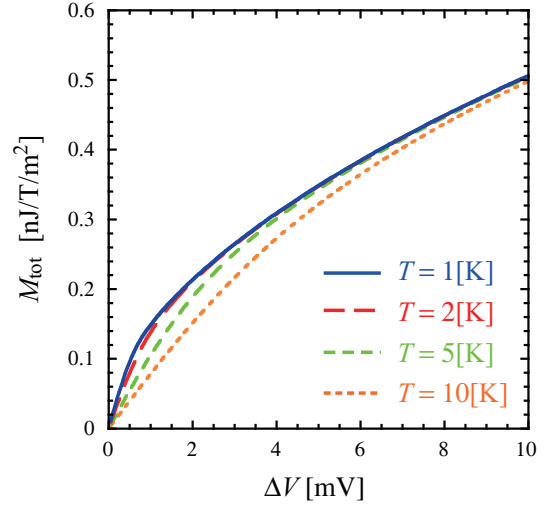}
\caption{(color online) The magnetization dependence on the voltage bias $\Delta V$ for $\mu=0$[meV].}
\label{fig_M_DeltaMu}
\end{figure}

Going back to Eq.~(\ref{eq180}), we show in Fig.~\ref{fig_M_DeltaMu} the dependence of the magnetization on the voltage bias $\Delta V=-|e|\Delta\mu$ beyond the linear response.
The magnetization grows monotonically in this regime.
Although the contribution from a channel saturates as the voltage bias exceeds the energy gap of the channel, more and more channels contribute to the summation in Eq.~(\ref{eq180}) and hence the monotonic increase.

\section{Numerical calculation for a ladder model}
\label{sec-numerical}

\begin{figure}
\centering
\includegraphics[width=0.8\columnwidth]{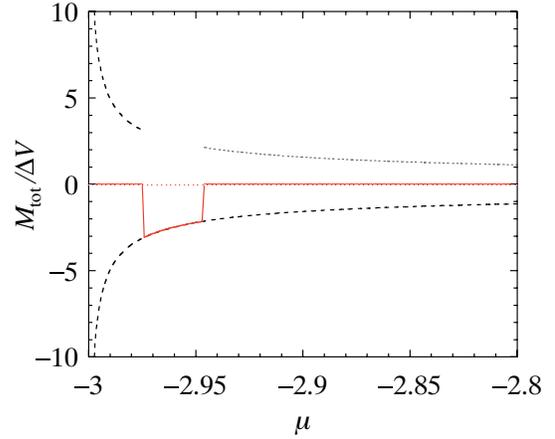}
\caption{(color online) The magnetization in the $y$ direction of the model~(\ref{eq410}) for $t=1$, $\theta a=0.2$ and $H_x=0.01$.
There are two subbands in this region, just as in Fig.~\ref{fig_channel}.
The broken curves in the negative side indicates the contribution of the right-going states with $k_x>0$ in the lower subband (the second integral of Eq.~(\ref{eq170})), the broken curve in the positive side indicates the contribution of the left-going states with $k_x>0$ in the lower subband (the first integral of Eq.~(\ref{eq170})), and the dotted curve indicates the contribution of the left-going states in the upper subband (the third integral of Eq.~(\ref{eq170})).
The solid curve is the sum of the three contributions, which survives only in the energy gap.}
\label{fig_scandisp}
\end{figure}

So far, we have assumed periodic boundary conditions in the $y$ direction.
When we introduce a confining potential $U(y)$ into the Hamiltonian~(\ref{eq10}), the gap at $k_x=0$ may be closed;
the standing wave in the $y$ direction includes both positive and negative $k_y$ and thus the gap-opening effect of the term $\hat{p}_y\hat{\sigma}_x$ is cancelled out.
We can then revive the gap by introducing a magnetic field in the $x$ direction: $-H_x\hat{\sigma}_x$.
In order to confirm our prediction in the confined geometry, we carried out a numerical calculation of the corresponding tight-binding model on a two-leg ladder~\cite{Ando92}:
\begin{eqnarray}
\label{eq410}
\mathcal{H}&=&-t\sum_{x=-\infty}^\infty
\sum_{\sigma=\uparrow,\downarrow}
\Biggl(
\sum_{y=1,2}
{c_{x+1,y,\sigma}}^\dag
e^{i\theta a\sigma_y}
c_{x,y,\sigma}
\nonumber\\
&&
\phantom{-t\sum_{x=-\infty}^\infty
\sum_{\sigma=\uparrow,\downarrow}}
+
{c_{x,2,\sigma}}^\dag
e^{-i\theta a\sigma_x}
c_{x,1,\sigma}
+\mathrm{H.c.}
\Biggr)
\nonumber\\
&&
-H_x\sum_{x=-\infty}^\infty\sum_{y=1,2}\sum_{\sigma=\uparrow,\downarrow}
{c_{x,y,\sigma}}^\dag \sigma_x c_{x,y,\sigma}
\end{eqnarray}
where $a$ is the lattice constant.
Figure~\ref{fig_scandisp} shows a result for $T=0$, confirming our prediction.
Note here that the width of the energy gap is of the order of $H_x$ but the magnitude of the magnetization is of the order of $\theta$.
Details of the calculations on the lattice system will be reported elsewhere.

The magnetic field $H_x$ applied in the present section in order to revive the energy gap at $k_x=0$ may induce additional spin polarization at the left and right ends of the system, namely at the contacts to the baths.
Even without the magnetic field, the dissipation at the contacts might generate some form of spin polarization.
The precise magnitude of the polarization would strongly depend on the interaction between the system and the baths and is beyond the scope of the present paper.
The magnitude, however, should be of the order of $\mathrm{O}(L^{-1})$ compared to the magnetization $M_\mathrm{tot}$;
the magnetization given in Eq.~(\ref{eq180}) and Fig.~\ref{fig_scandisp} is the one per unit area, that is, a bulk quantity.
We hence assume here that the spin polarization localized around the contacts is negligible.

Finally, the Hamiltonian with the Dresselhaus interaction takes the form~\cite{Rashba03}
\begin{equation}
\label{eq300}
\hat{\mathcal{H}}=\frac{
{\hat{p}_x}^2+{\hat{p}_y}^2
}{2m^\ast}
+\alpha_\mathrm{DSO} \left(
\hat{p}_x \hat{\sigma}_x
-\hat{p}_y \hat{\sigma}_y
\right).
\end{equation}
The rest of the formulation is the same as above with $\theta\equiv 2m^\ast\alpha_\mathrm{DSO}/\hbar$ and with the spin quantization in the $x$ direction as in
\begin{equation}
\label{eq320}
\hat{\sigma}_x=\left(
\begin{array}{cc}
1 & 0 \\
0 & -1
\end{array}
\right),
\quad
\label{eq330}
\hat{\sigma}_y=\left(
\begin{array}{cc}
0 & -i \\
i & 0
\end{array}
\right),
\quad
\label{eq340}
\hat{\sigma}_z=\left(
\begin{array}{cc}
0 & -1 \\
-1 & 0
\end{array}
\right).
\end{equation}
The magnetization in the $x$ direction is measured but the expression is the same as Eq.~(\ref{eq180}).

\section{Summary}

To summarize, we predicted that the magnetization appears under a voltage bias in pseudo-one-dimensional systems when the Fermi levels are tuned to be in the energy gap due to spin-orbit interactions.
The magnetization is an indication of the spin-carrying current, but its origin is essentially different from the one in two-dimensional systems with spin-orbit interactions.
The energy gap due to spin-orbit interactions plays a key role in exotic phenomena of pseudo-one-dimensional systems.
Without taking account of scattering processes, the argument may be naive;
we nevertheless believe that it is worth reporting and should be checked experimentally.
We can also argue that the magnetization appears under a temperature gradient~\cite{NHS05,SNH,NHS06} rather than a voltage gradient;
this effect will be reported elsewhere.

\section*{Acknowledgments}

The work is supported partly by Grand-in Aid for Exploratory Research (No.~17654073) from the Ministry of Education, Culture, Sports, Science and Technology and partly by the Murata Science Foundation as well as by the National Institutes of Natural Sciences undertaking Forming Bases for Interdisciplinary and International Research through Cooperation Across Fields of Study and Collaborative Research Program (No. NIFS06KDBT005).
One of the authors (N.H.) acknowledges support by Grant-in-Aid for Scientific Research (No.~17340115) from the Ministry of Education, Culture, Sports, Science and Technology as well as support by Core Research for Evolutional Science and Technology of Japan Science and Technology Agency.
The numerical integration in this work was done on a facility of the Supercomputer Center, Institute for Solid State Physics, University of Tokyo.

\end{document}